\documentstyle[12pt]{article}
\@addtoreset{equation}{section}
\def\theequation{\arabic{section}.\arabic{equation}}

\def\section{\@startsection{section}{1}{\z@}{3.5ex plus 1ex minus
   .2ex}{2.3ex plus .2ex}{\large\bf}}
\def\thesection{\Roman{section}.}

\def\appendix{\setcounter{section}{0}
        \def\thesection{Appendix\ \Alph{section}.}
        \def\theequation{\Alph{section}.\arabic{equation}}}

\newcommand{\beq}{\begin{eqnarray}}
\newcommand{\eeq}{\end{eqnarray}}
\newcommand{\eq}{eqnarray}

\newcommand{\al}{{\alpha}}
\newcommand{\be}{{\beta}}
\newcommand{\ci}{\cite}
\newcommand{\ga}{{\gamma}}

\newcommand{\ep}{{\epsilon}}

\newcommand{\de}{{\delta}}
\newcommand{\De}{\Delta}

\newcommand{\La}{{\Lambda}}
\newcommand{\m}{{\mu}}
\newcommand{\n}{{\nu}}

\newcommand{\om}{{\omega}}
\newcommand{\Om}{{\Omega}}
\newcommand{\pa}{{\partial}}
\newcommand{\no}{{\nonumber}}
\newcommand{\f}{\frac}
\newcommand{\ra}{\rightarrow}
\newcommand{\lra}{\leftrightarrow}
\newcommand{\eff}{\hbox{\scriptsize eff}}
\newcommand{\tot}{\hbox{\scriptsize tot}}
\newcommand{\stat}{\hbox{\scriptsize stat}}

\newcommand{\new}{\hbox{\scriptsize new}}

\newcommand{\BH}{\hbox{\scriptsize BH}}

\newcommand{\ther}{thermodynamics }
\newcommand{\therl}{thermodynamical }

\newcommand{\temp}{temperature }

\newcommand{\hb}{\hat{\be}}
\newcommand{\hO}{\hat{\Om}}
\newcommand{\he}{\hat{\eta}}

\newcommand{\na}{\nabla}
\begin{document}
\topmargin 0pt \oddsidemargin -3.5mm \headheight 0pt \topskip 0mm
\addtolength{\baselineskip}{0.20\baselineskip}
\begin{flushright}
   hep-th/0609027
\end{flushright}
\vspace{0.1cm}
\begin{center}
   {\large \bf BTZ Black Hole with Higher
  Derivatives, the Second Law of Thermodynamics,
   and Statistical Entropy: A New Proposal
  }
\end{center}
\vspace{0.1cm}
\begin{center}
 Mu-In Park\footnote{Electronic address:
muinpark@yahoo.com}
\\{Research Institute of Physics and Chemistry,}
\\ {Chonbuk National University, Chonju 561-756, Korea}
\end{center}
\vspace{0.1cm}
\begin{center}
  {\bf ABSTRACT}
\end{center}
I consider the thermodynamics of the BTZ black hole in the presence
of the higher curvature and gravitational Chern-Simons terms, and
its statistical entropy. I propose  a {\it new} thermodynamical
entropy, which being {\it non-negative} manifestly,  such as the
second law
of thermodynamics is
satisfied. I show that the new \therl entropy agrees perfectly with
the statistical entropy for
$all$ the values of the conformal factor of the higher curvature
terms and the coupling constant of the gravitational Chern-Simons
term, in contrast to some disagreements in the literatures. The
agreement with both the higher curvature and gravitational
Chern-Simons terms is possible because of an appropriate balancing
of them, though it is not a trivial matter because of a conflict in
the appropriate Hilbert space for a well-defined conformal field
theory for each term.
 \vspace{0.1cm}
 \vspace{5cm}
\begin{flushleft}
\\
21 April 2008 \\
\end{flushleft}
\newpage

\begin{section}
{Introduction}
\end{section}

Recently, the higher curvature corrections to the black hole
entropies of the black holes in supergravity theories in diverse
dimensions have been extensively studied, and it has been found that
there are some good agreements with the statistical entropies from
the microscopic counting of the number of states. ( For a recent
review, see Ref. \cite{Moha:05}. )

In all these analyses, the basic formalism for the \therl entropies
is known as the Wald's formalism which provides a general entropy
formula, based on the first law of thermodynamics, in the presence
{\it any covariant} combinations of the curvatures
\cite{Jaco:94,Iyer:94,Jaco:95}. But, there is a serious and
well-known problem in this formalism: The second law of \ther is
$not$ manifest, in contrast to the first law\footnote{For a class of
higher curvature theories where the Lagrangian is a polynomial in
the Ricci scalar $R$, the second law can be proved with the null
(matter) energy condition and cosmic censorship. But, for other
cases, there has been no general proof.}. However, it does not seem
that this question has been well explored in the recent studies of
supergravity black holes. Actually, without the guarantee of the
second law, there would be no justification for identifying the
entropies, even though they satisfy the first law \cite{Beke:73}.

More recently, the corrections due to the gravitational Chern-Simons
term \cite{Dese:82,Dese:91,Dese:02} have been studied in several
different approaches, and it has been found that there are good
agreements between the thermodynamical entropies based on the first
law, and the statistical entropies based on the boundary conformal
field theories (CFT) \cite{Krau:05b,Krau:05a,Solo:05a,Saho:06}. But,
the agreements were not perfect and there were some discrepancies in
a strong coupling regime, though not been well studied in the
literatures.

In order to resolve the discrepancies I have re-considered the first
law and argued that they can be resolved by considering a new
entropy formulae such as the second law
is guaranteed from some {\it new re-arrangements} of the usual form
of the first law \cite{Park:06,Park:06b}. In this paper, I study
$general$ higher curvature corrections as well and show that there
is similar discrepancies for a ``negative'' conformal factor ($\hO
<0$), in which the thermodynamic entropy become {\it negative},
which has been claimed ``unphysical'' in the literatures
\cite{Said:00} or speculated as an indication of some thermodynamic
instability \ci{Cvet:0112,Clun:04}. But, I argue that this can be
resolved also by considering appropriate new entropy formulae, which
being manifestly non-negative and satisfying the second law,
similarly to the case with the gravitational Chern-Simons. And, I
show that the new \therl entropy agrees perfectly with the
statistical entropy for
$all$ values of the conformal factor of the higher curvature terms
and the coupling constant of the gravitational Chern-Simons term.
Here, the agreement with both the higher curvature and gravitational
Chern-Simons corrections is not so trivial because the appropriate
Hilbert spaces for well-defined CFT are in conflict, but I find that
this is actually possible in our case, due to a nice balancing of
the two Hilbert spaces.

The plan of this paper is as follows.

In Sec. II, I consider the BTZ black hole in the presence of the
generic, higher curvature corrections, and I identify new entropies,
which being manifestly non-negative, such as the second law
of \ther can be satisfied. The obtained entropy
agrees with the usual Wald's formula for the positive conformal
factor $\hO>0$. However, it $disagrees$ with the Wald's formula for
$\hO <0$, which gives a {\it negative} entropy.

In Sec. III, I consider the statistical entropies and I find perfect
agreements with the new \therl entropies that have been found in
Sec. II, even for the $\hO<0$ case as well as the $\hO>0$ case.

In Sec. IV and V, I consider the gravitational Chern-Simons
correction term as well, in addition to the generic higher curvature
terms. I find perfect agreements between the \therl and statistical
entropies for
``all'' values (either $\hO>0$ or $\hO<0$) of the conformal factor
of the higher curvature terms and the coupling constant of the
gravitational Chern-Simons. The agreement with both of the two
corrections is possible because of an appropriate balancing of them,
though it is not a trivial matter because of a conflict in the
appropriate Hilbert space for a well-defined CFT for each
correction.

In Sec. VI, I conclude with several open questions.

In this paper, I shall keep the Newton's constant $G$ and the
Planck's constant $\hbar$ in order to clearly distinguish the
quantum gravity effects with the classical ones. But, I shall use
the units of $c\equiv 1,~k_B \equiv 1$, for the speed of light $c$
and the Boltzman's constant $k_B$ for convenience, as usual. \\

\begin{section}
{The BTZ black hole with higher curvatures}
\end{section}

The (2+1)-dimensional gravity with higher curvature terms and a
(bare) cosmological constant $\La=-1/{l_0}^2$ can be $generally$
described by the action on a manifold ${\cal M}$ [ omitting some
boundary terms ]
\begin{\eq}
\label{Higher} I_{g}=\frac{1}{16 \pi G} \int_{\cal M} d^3 x
\sqrt{-g} \left( ~f(g^{\m \n}, R_{\m \n}, \nabla_{\m}) +\frac{2}
{{l_0}^{2}}~\right),
\end{\eq}
where $f(g^{\m \n}, R_{\m \n}, \nabla_{\m})$ is an $arbitrary$
scalar function constructed from the metric $g^{\m \n}$, Ricci
curvature tensor $R_{ \m \n}$, and the covariant derivatives
$\nabla_{\m}$ \cite{Jaco:94,Iyer:94,Jaco:95}. This action is the
most generic, diffeomorphically invariant form in three dimensions
since there is no independent component of the Riemann tensor due to
vanishing Wely tensor. The equations of motion, by varying
(\ref{Higher}) with respect to the metric, are
\begin{\eq}
\label{eom:Higher}
 \f{\pa f}{\pa g_{\mu \nu}}-\f{1}{2} g^{\m \n} f -\f{1}{{l_0}^2} g^{\m \n}
=t^{\m \n},
\end{\eq}
where the pseudo-tensor $t^{\m \n}$ is given by
\begin{\eq}
t^{\m \n}=\f{1}{2} ( \na^\n \na^\al {P_\al}^\m + \na^\m \na^\al
{P_\al}^\n -\Box P^{\m \n}-g^{\m \n} \na^\al \na^\be P_{\al \be} )
\end{\eq}
with $P_{\al \be} \equiv g_{\al \m} g_{\be \n} (\pa f/\pa R_{\m
\n})$.

In the absence of the higher curvature terms, there is a black hole
solution, known as the BTZ ( Banados-Teitelboim-Zanelli ) solution,
which is given by the metric \cite{Bana:92}
\begin{eqnarray}
\label{BTZ}
 ds^2=-N^2 dt^2 +N^{-2} dr^2 +r^2 (d \phi +N^{\phi}
dt)^2
\end{eqnarray}
with
\begin{eqnarray}
\label{BTZ:N}
 N^2=\frac{(r^2-r_+^2) (r^2-r_-^2)}{{l_0}^2 r^2},~~
N^{\phi}=-\frac{r_+ r_-}{{l_0} r^2}.
\end{eqnarray}
Here, $r_+$ and $r_-$ denote the outer and inner horizons,
respectively. The mass and angular momentum of the black hole are
given by
\begin{eqnarray}
\label{mj:bare}
 m=\frac{r_+^2 +r_-^2}{8 G {l_0}^2},~~j=\frac{2 r_+ r_-}{8G l_0 },
\end{eqnarray}
respectively. Note that these parameters satisfy the usual
mass/angular momentum inequality $m^2 \geq j^2/{l_0}^2$ in order
that the horizon exists or the conical singularity is not naked,
with the equality for an extremal black hole having the overlapping
inner and outer horizons. This satisfies the usual Einstein equation
\begin{\eq}
\label{eom}
 R^{\mu \nu}-\f{1}{2} g^{\m \n} R -\f{1}{{l_0}^2} g^{\m \n}
=0
\end{\eq}
with a constant curvature scalar $R=-6/{l_0}^2$.

But, even in the presence of the generic higher curvature terms, the
BTZ solution can be still a solution since the $local$ structure
would be ``unchanged'' by higher curvatures. The only effects would
be some ``re-normalization'' of the bare parameters $l_0,r_+$, and
$r_-$ \cite{Krau:05b,Solo:05a,Saho:06,Said:00}: The renormalized
cosmological constant will be denoted by\footnote{In the
renormalized frame, one can also ``construct'' the Einstein equation
$R^{\mu \nu}-\f{1}{2} g^{\m \n} R -\f{1}{{l}^2} g^{\m \n} =0$, as in
(\ref{eom}), from the relations $R_{\m \n \al \be}=(R/6) (g_{\m
\al}g_{\n \be}- g_{\m \be}g_{\n \al})$ and $R=-6/l^2$.}
$\La_{ren}=-1/l^2$ and the function $l=l(l_0)$ depends on the
details of the function $f$\footnote{For $f=R+a R^2+b R_{\m \n}R^{\m
\n}$ with some appropriate coefficients $a,~b$ \cite{Said:00}, the
function $l=l(l_0)$ is given by $-6 l^{-2}=\left(-1 \pm \sqrt{1-24
(b-a)l_0^{-2}}\right)/(2 (b-a))$.}; however, I shall use the same
notations $r_\pm$ in the renormalized frame also, for brevity. And,
in this case one finds $t^{\m \n}=0$ trivially from $P_{\al
\be}\propto g_{\al \be}$ for any {\it constant-curvature} solution
\cite{Said:00}. On the other hand, the original mass and angular
momentum, computed from the standard Hamiltonian approach
\cite{Wald:93,Iyer:94}, become
\begin{\eq}
\label{MJ}
 M=\hO m,~~J=\hO j ,
\end{\eq}
respectively, where the conformal factor $\hO$ is defined by
\begin{\eq}
\label{Om}
 \hO \equiv \f{1}{3} g_{\m \n} \f{\pa f}{\pa R_{\m \n}},
\end{\eq}
which being constant for any constant-curvature solution
\cite{Said:00}.
Note that $\hO$ is ``not''
positive definite such as the usual inequality of the mass and
angular momentum would $not$ be valid generally\footnote{Here, $m$
and $j$ represent the usual mass and angular momentum for the metric
(\ref{BTZ}) in the {\it renormalized} frame $m=\frac{r_+^2 +r_-^2}{8
G {l}^2},~~j=\frac{2 r_+ r_-}{8G l }$, with the renormalized
parameters $l,r_\pm$, such as one has the usual inequality $m-j/l
\geq 0$ still.}
\begin{\eq}
\label{MJ:bound}
 M-J/l= \hO (m-j/l),
\end{\eq}
but it depends on the sign of $\hO$: For $\hO>0$ (case (i)) one has
the usual inequality $M \geq J/l$; however, for $\hO <0$ (case
(ii)), one has an {\it anomalous} inequality $J/l \geq M$ with the
``negative'' $M$ and $J$, though their magnitudes still satisfy the
usual bound $M^2 \geq J^2/l^2$.

Now, by considering the first law of \ther
\begin{eqnarray}
\label{first:old}
 \delta M=\Omega_+\delta J + T_+ \delta S_W,
\end{eqnarray}
with the Hawking temperature $T_+$ and the angular velocity
$\Omega_+$ of the (outer) event horizon $r_+$
\begin{eqnarray}
\label{Hawk:Temp}
 T_+=\left. \frac{\hbar \kappa}{2 \pi}
\right|_{r_+}=\frac{\hbar (r_+^2 -r_-^2)}{2 \pi l^2
r_+},~~\Omega_+=\left.-N^{\phi} \right|_{r_+}=\frac{r_-}{l r_+}
\end{eqnarray}
for the surface gravity function
\begin{\eq}
 \kappa=\f{1}{2} \f{\partial N^2}{
\partial r},
\end{\eq}
the black hole entropy can be identified as
\begin{eqnarray}
\label{S:Wald}
 S_W=\hO \frac{2 \pi r_+}{4 G \hbar}.
\end{eqnarray}
This agrees with the Wald's entropy formula \cite{Said:00}, and this
should be the case since the entropy in the Wald's formalism is
basically defined by the first law of \ther \cite{Jaco:95}. But, I
must note that the first law can not be ``proved'' without knowing
the form of the entropy and temperature, basically. Actually, in the
case of black holes, we usually know the Hawking temperature, as in
(\ref{Hawk:Temp}), from the Hawking radiation analysis with a given,
{\it Riemanian}\footnote{For some non-Riemanian geometry, it seems
that a different Hawking temperature could occur. See Ref.
\ci{Park:0610} for this possibility.}, metric and so we can identify
the entropy, by ``assuming'' the first law. This is a basic process
to compute the entropy in the general class of gravity theories
(see, for example, Refs. \cite{Iyer:94,Jaco:95}).

However, a basic problem of this approach is that the second law is
{\it not} guaranteed, in general. Actually, in the higher derivative
gravity theories there would be deep changes in the entropy, and the
second law or the Hawking's area (increasing) theorem has to be
revisited completely, generally. But, in regards to the area
theorem, this is ``not" true in our case: Our space is maximally
symmetric, i.e., a constant curvature space, and so the higher
curvature effects to the energy momentum tensor, $t_{\mu \nu}$ of
(2.3) vanishes, as I have clearly noted in the paragraph below
(2.7). Other higher curvature effects in the Einstein equation (2.2)
from the arbitrary function $f$, give only some re-normalization of
the bare parameters $l_{0},r_{+},r_{-}$, as I have explained in the
same paragraph, such as the resulting equation be just the original
vacuum Einstein equation (2.7) with the parameters'
re-normalization\footnote{There are an explicit comment about this
in the literature (see the footnote 2 in Ref. \ci{Saho:06}) and an
explicit computation also (see the Example in Sec.3.4 in Ref.
\ci{Said:00}); and see also the two footnotes 3 and 4, in this
paper, about this. This has been also discussed more explicitly in
Ref. \ci{Park:0610}.}. So, as far as the area theorem for the outer
horizons is concerned, the usual derivations via the Raychaudhuri's
equation with the null energy condition for matter's energy-momentum
tensor still works in our case since the $vacuum$ Einstein equation
satisfies the null-energy condition, trivially \ci{Park:0610}.
Moreover, there is another approach, called the ``physical process''
\ci{Wald:94,Jaco:95}), to prove the area theorem which does not
depends on the details of the gravity \ci{Jaco:94}. It depends only
on the first law with an appropriate energy condition, and the area
law is evident, in this approach, from the first law (2.11) also.
(For the details, see Ref. \ci{Park:0611}.) Hence, it is clear that
the entropy formula (\ref{S:Wald}) satisfies the second law, for the
case of $\hO
>0$, since it is already in the  Bekenstein's form, which being
proportional to the area of the outer horizon ${\cal A}_+=2 \pi
r_+$, such as the Hawking's area (increasing) theorem implies the
increasing entropy \cite{Hawk:71}.

On the other hand, the situation is quite different for
$\hO<0$\footnote{Note that our theory with the negative conformal
factor does not affect the causal strcture in the Einstein frame
either since the (three-dimensional) frame transformation
$\bar{g}_{\m \n}=\hat{\Om}^2 g_{\m \n}$ is insensitive to the sign
of $\hat{\Om}$ \ci{Jaco:95,Said:00}.}
%
since (\ref{S:Wald}) would {\it not} guarantee
the second law nor the {\it positiveness} because it would
``decrease'' indefinitely, with the negative values, as the outer
horizon $r_+$ be increased, from the area theorem. Actually, this
seems to be general feature of higher derivative gravities in
arbitrary dimensions \ci{Cai:0109,Cvet:0112,Cai:03,Clun:04} or {\it
Taub-Bolt} spacetime with a cosmological constant
\ci{Clun:04,Mann:99}, and in the literatures it has been speculated
as an indication of some thermodynamic instability (for example, see
Refs. \ci{Cvet:0112,Clun:04}). But, this seems to be physically
nonsensical since the entropy is {\it non-negative}, ``by its
definition'' as a measure of disorderedness \ci{Kitt:67}; the
positiveness of the entropy is a ``minimum'' requirement that must
be satisfied if the entropy has a statistical mechanical origin
\ci{Jaco:95}. Moreover, without the guarantee of the second law,
there would be no justification for identifying the entropies, even
though they satisfy the first law \ci{Beke:73}. So, in this paper I
consider a {\it different} approach which can resolve the two
problems, simultaneously. The new resolution is to consider an
entropy
\begin{\eq}
\label{S:W'} {S_W}' =|\hO | \frac{2 \pi r_+}{4 G \hbar},
\end{\eq}
which is non-negative manifestly and also satisfying the second law
from the area theorem\footnote{The physical process version of the
second law for this definition or the area theorem is related to the
same geometric effect as that of the $\hO>0$ case. (See Ref.
\ci{Park:0611} for the details.) }, as in the case of $S_W$ in
(\ref{S:Wald}) for a positive $\hO$.
But, in this
case I must pay the price, by considering a new temperature
\begin{\eq}
{T_+}' &\equiv& -T_+,
\end{\eq}
instead of $T_+$, in order to satisfy the first law also.

In the following sections, I will show that the new approach is
actually what favored by the statistical entropy through a CFT
analysis, which provides the new entropy formula (\ref{S:W'})
directly: With the correct values of $M,~J$, and the entropy
${S_{W}}'$, which is non-negative and proportional to the (outer)
horizon area, there is no other choice
in the temperature. \\


\begin{section}
{Statistical entropy}
\end{section}

It is well known that, in the absence of the higher curvature terms,
the statistical entropy of the BTZ black hole can be computed from a
two-dimensional CFT, which is described by Virasoro algebras, on the
asymptotic Anti-de Sitter ($AdS$) boundary with the help of the
Cardy formula, and there is a complete agreement with the \therl
Bekenstein-Hawking entropy. There are basically two approaches to
compute the CFT, i.e., the Virasoro algebras. One approach is a
$quantum$ approach which identifies the central charges of the CFT,
in the context of the $conjectured$ AdS/CFT correspondence
\cite{Ahar:00}, by evaluating the anomalies of the CFT effective
action on the AdS $boundary$, from the regularized $bulk$ gravity
action \cite{Henn:98,Hyun:99,Bala:99}. The other approach is a
classical one which $directtly$ computes (classical) Virasoro
algebras based on the classical symmetry algebras of the asymptotic
isometry of $AdS_3$ \cite{Brow:86,Bana:99,Oh:98,Park:98,Park:99}. It
is widely known that these two approaches agree completely, and this
provides an explicit check of the AdS/CFT correspondence.

Recently, these analyses have been generalized to the theories with
higher curvature terms, and some good agreements were known between
the \therl and statistical entropies, as well as the agreements
between the holographic anomaly approach and classical symmetry
algebra approach. But, in contrast to the usual claims in the
literatures, these analyses have some problems which ``might''
invalidate the AdS/CFT correspondence. First, there are some
discrepancies in the usual thermodynamical and statistical
entropies, though this has not been well explored in the literatures
\cite{Said:00,Krau:05b,Saho:06}. Second, the computation about the
classical symmetry algebra, by transforming a gravity action with
the higher curvature terms into the usual Einstein-Hilbert action
with some auxiliary {\it tensor matter} fields, ``might'' have some
problems since there would be some non-trivial, boundary
contributions in the Virasoro generators $\hat{L}^{\pm}_m$ and
central charges from the matter fields ``in general''
\cite{Park:04a,Henn:02,Gege:03}, in contrast to the work
\cite{Said:00}, though the agreements seems to be plausible in the
context of AdS/CFT. In this paper, the second problem will not be
discussed further and left as an open problem. In the remainder of
this paper, I will concentrate only the first problem and in the
context of the quantum approach of the ``holographic anomalies'',
which has been computed rigorously recently, such as the second
problem does not occur, in contrast to Ref. \cite{Said:00}.

Now, in order to discuss the first problem in detail, I start by
noting the holographic (conformal) anomalies in the expectation
values of the boundary stress tensor \cite{Krau:05b}, for a boundary
metric $ds^2 \simeq -r^2 dx^{+} dx^{-}$ with $r$ taken to infinity,
\begin{\eq}
\label{anomaly}
\left< T_{++} (x^+)\right>=-\f{\hbar \hat{c}^+}{24
\pi},~~\left< T_{--} (x^+)\right>=-\f{\hbar \hat{c}^-}{24 \pi}
\end{\eq}
with the central charges \cite{Noji:99,Krau:05b}[ I follow the
conventions of Ref. \cite{Bala:99} ]
\begin{\eq}
\hat{c}^{\pm}=\hO \frac{3 l}{2G \hbar}.
 \label{c:anomaly(high)}
\end{\eq}
Note that the obtained central charges have a {\it quantum} origin,
which would has been introduced via the regularization procedure.

By considering (\ref{anomaly}) as the anomalous transformations of
the boundary stress tensors under the diffeomorphism $\de
x^{\pm}=-\xi^{\pm}(x^{\pm})$,
\begin{\eq}
\de_{\xi^{+}}T_{++}&=&2 \pa_+ \xi^+ T_{++} +\xi^+ \pa_+
T_{++}-\f{\hbar \hat{c}^+}{24 \pi} \pa^3_+ \xi^+ \no \\
&=&\f{1}{i}[T_{++}, \hat{L}^+[\xi^+]], \no \\
 \de_{\xi^{-}}T_{--}&=&2 \pa_- \xi^-
T_{--} +\xi^- \pa_- T_{--}-\f{\hbar \hat{c}^-}{24 \pi} \pa^3_- \xi^-
\no \\
&=&\f{1}{i}[T_{--}, \hat{L}^-[\xi^-]]
\end{\eq}
with the generators
\begin{\eq}
\hat{L}^{\pm}[\xi^{\pm}]=\f{1}{\hbar} \oint dx^{\pm} T_{\pm \pm}
\xi^{\pm} (x^{\pm})+\f{\hat{c}^{\pm}}{24 },
\end{\eq}
one can obtain a pair of quantum Virasoro algebras
\begin{\eq}
[\hat{L}^{\pm}_{m},\hat{L}^{\pm}_{n}]=(m-n) \hat{L}^{\pm}_{m+n}+\f{
\hat{c}^{\pm}}{12 } m (m^2-1) \de_{m+n,0}
\end{\eq}
with the central charges $\hat{c}^{\pm}$ for the
right(+)/left(-)-moving sectors and for a monochromatic basis
$\xi^{\pm}=e^{imx^{\pm}}$ with the integer numbers $m$ and $n$. In
the absence of the higher curvature terms, this reduces to the usual
result for the holographic {\it conformal} anomaly of $AdS_3$ from
$\hO=1$ \cite{Henn:98,Hyun:99,Bala:99}, whereas higher curvature
terms produce the departures from the unity, i.e., $\hO-1$, which
can be either positive or negative, depending on the coupling
constants of the higher curvature terms.

Then, let me consider the ground state Virasoro generators,
expressed in terms of the black hole's mass and angular momentum:
\begin{\eq}
\label{ground}
 \hat{L}^{\pm}_{0}&=&\frac{lM \pm J}{2\hbar} +\f{
\hat{c}^{\pm}}{24
}\no \\
&=&\hO^{\pm} \frac{(lm \pm j)}{2\hbar} +\f{ \hat{c}^{\pm}}{24 }.
\end{\eq}

With the Virasoro algebras of $\hat{L}^{\pm}_{m}$ in the standard
form, which are defined on the {\it plane}, one can use the Cardy
formula for the asymptotic states
\cite{Card:86,Dijk:00,Park:02,Kang:04,Park:04b}
\begin{eqnarray}
\label{Cardy}
 \mbox{log}~ \rho (\hat{\Delta}^{\pm}) \simeq 2 \pi
\sqrt{ \f{1}{6} \left(\hat{c}^{\pm}-24
  \hat{\Delta}^{\pm}_{\hbox{\scriptsize
  min}}\right)\left(\hat{\Delta}^{\pm}-\frac{\hat{c}^{\pm}}{24}\right) },
\end{eqnarray}
where $\hat{\Delta}^{\pm}$ are the eigenvalues, called conformal
weights, of the operator $\hat{L}_0$ for black-hole quantum states
$|\hat{\Delta}^{\pm}\rangle$ and
$\hat{\Delta}^{\pm}_{\hbox{\scriptsize min}}$ are their minimum
values. Here, I note that the above Cardy formula, which comes from
the saddle-point approximation of the CFT partition function on a
torus, is valid only if the following two conditions are satisfied:
\begin{\eq}
\label{cond1}
\f{24 {\hat \De}^{\pm}_{\eff}}{\hat{c}^{\pm}_{\eff}} &\gg& 1, \\
\label{cond2}
 \hat{c}^{\pm}_{\eff} {\hat \De}^{\pm}_{\eff} &\gg& 1,
\end{\eq}
where $\hat{\Delta}^{\pm}_{\hbox{\scriptsize
eff}}=\hat{\Delta}^{\pm}
-\hat{c}^{\pm}/{24},~\hat{c}^{\pm}_{\hbox{\scriptsize
eff}}=\hat{c}^{\pm}-24\hat{\Delta}^{\pm}_{\hbox{\scriptsize min}}$
are the effective conformal weights and central charges,
respectively; from the first condition, the higher-order correction
terms are exponentially suppressed as $e^{- 2 \pi \ep^{\pm}
(\hat{\Delta}^{\pm}-\hat{\Delta}^{\pm}_{\hbox{\scriptsize min}})}$
with $\ep^{\pm} \equiv 24 {\hat
\De}^{\pm}_{\eff}/\hat{c}^{\pm}_{\eff}$; from the second condition,
the usual saddle-point approximation is reliable, i.e.,
$\rho(\hat{\Delta}^{\pm})$ dominates in the partition function (see
Refs. \cite{Park:04b,Park:06b} for the details).

Then, the statistical entropy for the asymptotic states becomes
\begin{\eq}
S_{\stat}&=&\mbox{log}~ \rho (\hat{\Delta}^{+}_0)+\mbox{log}~ \rho
(\hat{\Delta}^{-}_0) \no \\
&=&\f{\pi}{4 G \hbar} |\hO (r_+ + r_-)|+\f{\pi}{4 G \hbar} |\hO
(r_+ - r_-)| \no \\
&=&|\hO|~ \f{2 \pi r_+}{4 G \hbar} \label{S:stat}
\end{\eq}
where I have chosen $\hat{\Delta}^{\pm}_{0~ (\hbox{\scriptsize
min})}=0$ as usual \cite{Stro:98,Park:04a}; from (\ref{ground}),
this corresponds to the $AdS_3$ vacuum solution with $m=-1/(8G)$ and
$ j=0$ in the usual context, but one has
\begin{\eq}
M=-\f{\hO}{8G},~~J=0
\end{\eq}
in the new context. Note that the correct ``$1/\hbar$'' factor for
the semiclassical black hole entropy comes from the appropriate
recovering of $\hbar$ in (\ref{c:anomaly(high)}) and (\ref{ground}).
According to the conditions of validity (\ref{cond1}) and
(\ref{cond2}), this entropy formula is valid only when both of the
two conditions
\begin{\eq}
\label{cond3}
&&(r_+\pm r_-) \gg l, \\
\label{cond4}
 &&(r_+ \pm r_-) \gg \hbar G
\end{\eq}
are satisfied. The usual semiclassical limit of large black hole
(area), in which the back-reaction of the emitted radiation from the
black hole is neglected \cite{Pres:91,Cruz:04} and so the
thermodynamical entropy formula (\ref{S:Wald}) and (\ref{S:W'}) from
the first law can be reliable, agrees with the condition
(\ref{cond4}). The condition (\ref{cond3}) provides  one more
restriction on the black hole systems, though this does not seem to
be needed, in general. But, at this stage, the condition of large
central charges $\hat{c}^{\pm} \gg 1$, i.e., $l \gg \hbar G$
\cite{Stro:98}, which would be related to the leading supergravity
approximation of AdS/CFT correspondence \cite{Ahar:00}, is not
needed yet. It is interesting to note also that the statistical
entropy (\ref{S:stat}) from the Cardy formula (\ref{Cardy}) has
basically the {\it same form} for both the Einstein-Hilbert action
and the higher curvature corrected action; the only changes are some
correction terms in the central charges and the conformal weights
themselves, rather than considering the higher order corrections to
the Cardy formula itself ! This is because the higher curvature
terms do {\it not necessarily} imply the quantum corrections, such
as even the higher curvature terms can be treated semiclassically by
neglecting the back reaction effects, which are quantum effects, and
so (\ref{cond2}) or (\ref{cond4}) can be satisfied \cite{Park:06b}.

Now, let me consider the following two cases, depending on the signs
of $\hO$: (i). $\hO \geq 0$ and (ii). $\hO <0$.\\

(i). $\hO \geq 0$: In this case, I have $|\hO|=\hO$ and the
statistical entropy (\ref{S:stat}) becomes
\begin{eqnarray}
\label{S_stat:old2}
 S_{\stat}=\hO \frac{2 \pi r_+}{4 G \hbar}.
\end{eqnarray}
This agrees exactly with the usual Wald's entropy formula
(\ref{S:Wald}), as was known also in the literatures
\cite{Krau:05b,Saho:06,Said:00}. And, this is the case where
$\hat{c}^{\pm}$ and $\hat{\Delta}^{\pm} -\hat{c}^{\pm}/{24}$ are
positive definite such as the Cardy formula (\ref{Cardy}) has a
well-defined meaning. In the gravity side also it shows the usual
behavior with the ``positive'' mass and angular momentum satisfying
the normal inequality $M \ge J/l$, as well as $M^2 \ge J^2/l^2$.\\

(ii). $\hO <0$: In this case, I have $|\hO|=-\hO$ and so the
statistical entropy (\ref{S:stat}) becomes
\begin{eqnarray}
\label{S_stat:new}
 S_{\stat}=-\hO \frac{2 \pi r_+}{4 G \hbar}.
\end{eqnarray}
This agrees exactly with the {\it modified} entropy formula
(\ref{S:W'}), which is manifestly {\it positive} and guarantees the
second law of thermodynamics, even in this case. But, this can not
agree with the usual Wald's formula (\ref{S:Wald}), giving a {\it
negative} entropy, though this has $not$ been well-explored in the
literatures\footnote{The errors in Ref. \ci{Said:00} came from the
missing of ``absolute values'' in the computation; for example, in
(32) or (39), $\Omega$ should be corrected due to $\sqrt{\Omega^2}=
|\Omega|$. In other literatures \ci{Krau:05b,Krau:05a,Saho:06},
$\hat{\Omega}>0$ has been implicitly, or explicitly assumed,
instead.};
this discrepancy can be only resolved in the modified entropy
formula (\ref{S:W'}). And, this is the case where there is an
abnormal mass bound due to $M\le J/l $ with {\it negative} $M$ and
$J$. Moreover, in the CFT side also, this is not the usual system
either because $\hat{c}^{\pm}=\hO (3 l/2G \hbar)$ and
$\hat{\Delta}^{\pm} -\hat{c}^{\pm}/{24}=\hO (ml -j)/2\hbar$ are
$negative$ valued, but the CFT is perfectly well defined. The
application of the Cardy formula to the case of negative
$\hat{c}^{\pm}$ and $\hat{\Delta}^{\pm} -\hat{c}^{\pm}/{24}$ might
be questioned due to the existence of negatives-norm states with the
usual condition $\left.
\hat{L}^\pm_n|\hat{\Delta}^{\pm}\right>=0~(n>0)$ for the
highest-weight state $\left.|\hat{\Delta}^{\pm}\right>$. However,
this problem can be easily cured, though not quite well-known, by
considering another representation of the Virasoro algebras with
$\tilde{L}_{n}^\pm\equiv
-\hat{L}_{-n}^\pm,~\tilde{c}^\pm\equiv-\hat{c}^\pm$ and
$\tilde{L}_n^\pm|\tilde{\De}^\pm \left.\right>=0~(n>0)$ for the new
highest-weight state $|\tilde{\De}^\pm\left.\right>$
\cite{Bana:99b,Park:06,Park:06b}; this implies that the Hilbert
space need to be ``twisted'' in which the whole states vectors be
constructed from the {\it doubly}-twisted highest-weight state
$\left.|\tilde{\Delta}^{+}\right>
\otimes\left.|\tilde{\De}^-\right>$. The formula (\ref{S:stat}),
which is invariant under this substitution--actually their
self-compensations of the negative signs produce the $real$ and
$positive$ statistical entropy, should be understood in this
context.

In summary, I have found exact agreements between the {\it new}
thermodynamic black hole entropies which are manifestly
non-negative, satisfying the second law, and have been evaluated in
the bulk (AdS) gravity side and the CFT entropies in the asymptotic
boundary, for
 $any$ value of the conformal factor $\hO$ of higher curvature
gravities. So, the modified entropy formula for a negative conformal
factor $\hO <0$
seems to be supported by the
sub-leading order with generic higher curvature terms, as well as in
the leading order with
the Einstein-Hilbert action.\\

\begin{section}
{Inclusion of a gravitational Chern-Simons term (I): Thermodynamics}
\end{section}

In three ( or odd in general ) dimensions, the gravitational
Chern-Simons term can also be included as a higher {\it derivative}
correction, as well as higher curvature corrections. The total
action with a gravitational Chern-Simons term as well as generic
higher curvature terms is described  by the action
\begin{\eq}
\label{I:tot}
 I_{g(\tot)}=I_g +I_{GCS},
\end{\eq}
where the gravitational Chern-Simons term is given by\footnote{ Note
that the $dimensioless$ coupling constant $\hat{\beta}$ is related
to the one used in Refs. \cite{Dese:82,Dese:91,Dese:02} as
$\hat{\beta}=-1/( \mu l_0)$, in Ref. \cite{Solo:05a} as
$\hat{\beta}=-\beta_S/l_0$, and in Ref. \cite{Krau:05a} as
$\hat{\beta}=-32 \pi G \beta_{KL}/l_0$. } [ the Greek letters
($\mu,\nu,\alpha, \cdots$) denote the space-time indices, and the
Latin ($a,b,c, \cdots$) denote the internal Lorentz indices; I take
the metric convention $\eta_{ab}$=diag$(-1,1,1)$ for the internal
Lorentz indices, and the indices are raised and lowered by the
metric $\eta_{ab}$ ]
\begin{eqnarray}
\label{GCS}
 I_{GCS}=\frac{\hat{\beta} l_0}{64 \pi G} \int_{\cal M} d^3 x
~\epsilon^{\mu \nu \alpha} \left( R_{ab \mu \nu}
{\omega^{ab}}_\alpha +\frac{2}{3}~ {\omega^b}_{c_\mu} {\omega^c}_{a
\nu} {\omega^a}_{b\alpha} \right).
\end{eqnarray}
Here, the spin-connection $1-form$ ${\omega^a}_b={\omega^a}_{b \mu}
dx^{\mu},~\omega_{ab\mu}=-\omega_{ba\mu}$ is determined by the
torsion-free condition
$
 d e^a +{\omega^a}_b \wedge e^b=0
$ with the dreibeins 1-form $e^a={e^a}_{\mu} dx^{\mu}$, and the
curvature is then $R_{ab \mu \nu}=\pa_{\mu}\omega_{ab
\nu}+{{\om_a}^c}_{\mu} \om_{cb\nu}-(\m \lra \n)$. [ I take the same
definitions as in Ref. \cite{Krau:05a} for the curvature 2-form
$R_{ab}=(1/2)R_{ab\m \n}~dx^{\m} \wedge dx^{\n}$ and the
spin-connection 1-form $\om_{ab}$.] Note that $I_{GCS}$ is of third
derivative order and it does not have the diffeomorphism symmetry in
the ``bulk''.

The resulting equations of motion are
\begin{\eq}
 \f{\pa f}{\pa g_{\mu \nu}}-\f{1}{2} g^{\m \n} f -\f{1}{{l_0}^2} g^{\m \n}
=t^{\m \n}+\hat{\be} l_0 C^{\m \n},
\end{\eq}
where the Cotton tensor $C^{\m \n}$ is defined by
\begin{\eq}
C^{\mu \nu}= \f{1}{\sqrt{-g}} \epsilon^{\mu \rho \sigma}
\nabla_{\rho} ({R^{\nu}}_{\sigma}-\f{1}{4} {\delta^{\nu}}_{\sigma} R
),
\end{\eq}
which is traceless and covariantly conserved \cite{Dese:82}. The BTZ
solution (\ref{BTZ}), (\ref{BTZ:N}) satisfies the same equations of
motion as (\ref{eom}) from $C^{\m \n}=0$, like as $t^{\m \n}=0$, for
any constant curvature solution, and of course with the renormalized
parameters $l,r_+$, and $r_-$.

In the absence of the higher curvature terms, the mass and angular
momentum are found to be\footnote{ This has been checked in several
different approaches, e.g., the quasi-local method's in Ref.
\cite{Garc:03}, the super-angular momentum's in Ref. \cite{Mous:03},
the ADM's in Refs. \cite{Dese:05,Olme:05}, the holography's in Refs.
\cite{Krau:05a,Solo:05a}.}
\begin{eqnarray}
 M_{GCS}=m+\hat{\be} j/l_0, ~~ J_{GCS}=j+ \hat{\be}l_0 m.
 \label{MJ:GCS}
\end{eqnarray}
So, in the presence of the higher curvature terms as well as the
gravitational Chern-Simons term, one can evaluate the {\it total}
mass and angular momentum as
\begin{\eq}
\label{M:tot}
 M_{\tot}&=&m+(\hO-1)m+\hat{\be} j/l, \no \\
&=&\hO m+\hat{\be} j/l, \\
\label{J:tot}
 J_{\tot}&=&j+(\hO-1)j+ \hat{\be}l m, \no \\
 &=&\hO j+ \hat{\be}l m,
\end{\eq}
by summing the two contributions with the appropriate
renormalization of the parameters $l_0,r_+,r_-$ \ci{Saho:06}. Here,
$(\hO-1)$-factors came from the higher curvature corrections of
(\ref{MJ}) and $\hb$-factors came from the gravitational
Chern-Simons corrections of (\ref{MJ:GCS}). In contrast to the case
with the higher curvature corrections only, the usual inequalities
of the mass and angular momentum are not generally valid,
\begin{\eq}
\label{M_bound:tot} M_{\tot} -J_{\tot}/l &=&\hO
\left(1-\f{\hb}{\hO}\right)
(m-j/l), \no \\
 M_{\tot}^2 -J_{\tot}^2/l^2 &=&\hO^2 \left(1-\f{\hb^2}{\hO^2}\right) (m^2-j^2/l^2)
\end{\eq}
but depends on  the values of the ratio, $\hat{\eta} \equiv
\hat{\be}/\hO$: For small values of ratio, $|\eta|<1$, the usual
inequality in magnitudes is preserved, i.e., $M_{\tot}^2 \geq
J_{\tot}^2/l^2$; however, for the large values of ratio, $|\eta|>1$,
one has an {\it anomalous} inequality with an exchanged role of the
mass and angular momentum as $J_{\tot}^2/l^2 \geq M_{\tot}^2$; also,
at the critical value $|\eta|=1$, the modified mass and angular
momentum are ``always'' saturated, i.e., $M_{\tot}^2 =
J_{\tot}^2/l^2$, regardless of inequality of the bare parameters
$m$,$j$ and the signs of $\hO$. But, the inequality for $M_{\tot}$
and $J_{\tot}$ depends on the sign of $\hO$, also.

Now, by considering the first law of thermodynamics as
\begin{eqnarray}
\label{first:old(tot)}
 \delta M_{\tot}=\Omega_+\delta J_{\tot} + T_+ \delta S_{W(\tot)}
\end{eqnarray}
with the temperature $T_+$ and angular velocity $\Omega_+$ of the
outer horizon $r_+$, the total entropy can be identified as
\begin{eqnarray}
\label{S:Wald(tot)}
 S_{W(\tot)}=\hO \left(~\frac{2 \pi r_+}{4 G \hbar}+\hat{\eta} \frac{2 \pi r_-}{4
G \hbar}~\right).
\end{eqnarray}
This agrees with the Wald's entropy formula \cite{Saho:06}, as it
should be.
But, as I have argued for the gravitational Chern-Simons corrected
case in the recent works \cite{Park:06,Park:06b} and in Sec. II for
the higher- curvature corrected case, the positiveness nor the
second law of \ther would not be guaranteed by the entropy
(\ref{S:Wald(tot)}), generally. Especially for the gravitational
Chern-Simons correction term in (\ref{S:Wald(tot)}), being
proportional to the inner-horizon area ${\cal A}_-=2 \pi r_-$, there
is no guarantee of the second law due to lack of area (increasing)
theorem for the inner horizon; rather, it seems like that this would
rather {\it decrease} due to the instability of the inner horizon
\cite{Stei:94,Bala:04}. The only way of guaranteeing the second law
from the entropy (\ref{S:Wald(tot)}) is to consider an appropriate
balancing of $\hO$ and $\hb$ such as the contributions from the area
of the outer-horizon area ${\cal A}_+=2 \pi r_+$ dominate those from
${\cal A}_-$: Since $r_+\geq r_-$ is always satisfied, this
condition is equivalent to $|\he|<1$ with $\hO >0$. Actually, this
is the case where the usual mass/angular momentum inequalities hold
from (\ref{M:tot})$\sim$(\ref{M_bound:tot}) and the system behaves
as an ordinary BTZ black hole, though there are some shifts and
conformal factor corrections in the mass, angular momentum, and
entropy \cite{Saho:06}.

On the other hand, for the other values of $\hO$ and $\he$, the
entropy of (\ref{S:Wald(tot)}) does not guarantee the positiveness
nor the second law, and I need some different forms of the entropy.
There are, including the above ordinary case, totally $2\times 3=6$
possible cases from $2$ possibilities for $\hO$ ( $\hO>0$, $\hO<0$ )
and $3$ possibilities for $\he$ ( $|\he| \leq 1,~\eta >1,~\eta <-1$
). Let me consider the following five cases, in addition to the case
of (a). $\hO>0$, $|\he | \leq 1$ for the ordinary black holes above,
depending on the values of $\hO$ and $\he$: (b). $\hO>0$, $\he > 1$,
(c). $\hO>0$, $\he < -1$, (d). $\hO<0$, $|\he | \leq 1$, (e).
$\hO<0$, $\he  > 1$, and (f).
$\hO<0$, $\he  <-1$.\\

(b). $\hO>0$, $\he > 1$: In order to study this case, I first note
the following identity in the BTZ system \cite{Park:06,Park:06b}
\begin{\eq}
\label{id1}
\de m&=&\Om_+ \de j +T_+ \de S_{BH} \\
\label{id2} &=&\Om_- \de j +T_- \de S_{-}
\end{\eq}
with the temperature $T_-$ and angular velocity $\Om_-$ of the inner
horizon $r_-$
\begin{eqnarray}
\label{T-} T_-=\left. \frac{\hbar \kappa}{2 \pi}
\right|_{r_-}=\frac{\hbar (r_-^2 -r_+^2)}{2 \pi l^2 r_-},~~
\Omega_-=\left.-N^{\phi} \right|_{r_-}=\frac{r_+}{l r_-}
\end{eqnarray}
and the usual Bekenstein-Hawking entropy
\begin{\eq}
S_{\BH}=\f{2 \pi r_+}{4 G \hbar}.
\end{\eq}
Here, the physical relevances of the parameters $T_-$ and $\Om_-$
are not clear. But, here and below, I use $T_-$, $\Om_-$ just for
convenience in identifying the new entropy, from the ``assumed''
first law of thermodynamics (\ref{id2}).\footnote{I have used the
definition of $\kappa$ as $\nabla ^{\nu} (\chi ^{\mu} \chi_{\mu}
)=-\kappa \chi^{\nu}$ for the horizon Killing vector $\chi^{\mu}$ in
order to determine its {\it sign}, as well as its magnitude. }

Now, let me consider, from (\ref{M:tot}) and (\ref{J:tot}),
\begin{\eq}
\label{first:new1}
 \de M_{\tot}- \Om_{-} \de J_{\tot} =\hO \left[\de m- \Om_{-} \de j +
\he( \de j/l-\Om_{-} \de m)\right],
\end{\eq}
instead of $\de M_{\tot}- \Om_{+} \de J_{\tot}$ in
(\ref{first:old(tot)}). Then, it is easy to see that the first two
terms in the right hand side become $T_{-} \de S_{-}$ by using the
second identity (\ref{id2}). And also, the final two terms in the
bracket become $T_{-} \de S_{BH}$ by using the first identity
(\ref{id1}) and another identity
\begin{\eq}
\label{Om-+} \Om_{-}=\Om_{+}^{-1} l^{-2}.
\end{\eq}
So, finally I find that (\ref{first:new1}) becomes a new
re-arrangement of the first law as
\begin{eqnarray}
\label{first:new2}
 \delta M_{\tot}=\Omega_-\delta J_{\tot} + T_- \delta S_{\new}
\end{eqnarray}
with a {\it new} black hole entropy
\begin{eqnarray}
S_{\new}=\hO \left(~\frac{2 \pi r_-}{4 G \hbar}+\he \frac{2 \pi
r_+}{4 G \hbar} ~\right). \label{S:new}
\end{eqnarray}
With the new entropy formula, it is easy to see the previous
argument for the second law of thermodynamics of (\ref{S:Wald(tot)})
in the small values of coupling as $|\he| <1$, with $\hO>0$, can now
be applied to that of (\ref{S:new}) in the large values of coupling
as $|\he| > 1$.\\

(c). $\hO>0$, $\he < -1$: In this case, the entropy formula
(\ref{S:new}) would $not$ guarantee the second law of thermodynamics
$nor$ the $positiveness$ of the entropy: The entropy would
``decrease'' indefinitely, with the negative values, as the outer
horizon $r_+$ be increased from the Hawking's area theorem
\cite{Hawk:71}. But, there is a simple way of resolution from the
new form of the first law (\ref{first:new2}), as in the case of
$\hO<0$ with the higher curvature terms only in Sec. II. It is to
consider
\begin{eqnarray}
{S_{\new}} '&\equiv& -S_{\new}=-\hO \left(~\frac{2 \pi r_-}{4 G
\hbar}+\he \frac{2
\pi r_+}{4 G \hbar} ~\right), \label{S:new'} \\
\label{T-'}
 {T_-}' &\equiv& -T_-=\frac{\hbar (r_+^2 -r_-^2)}{2 \pi
l^2 r_-},
\end{eqnarray}
instead of $S_{\new},T_-$ and actually this choice seems to be
$unique$: One might consider ${S_{\new}} ''\equiv \hO \left[\frac{2
\pi r_-}{4 G \hbar}-\he \frac{2 \pi r_+}{4 G \hbar}\right]$, but
then the first law (\ref{first:new2}) is $not$ satisfied.\\

(d). $\hO<0$, $|\he | \leq 1$: This is similar to the case (ii) of
Sec. II and III such as the appropriate entropy formula is
\begin{\eq}
\label{S:W(tot)'}
 {S_{W(\tot)}}'=-S_{W(\tot)}=-\hO \left(~\frac{2 \pi r_+}{4 G
\hbar}+\he \frac{2 \pi r_-}{4 G \hbar} ~\right),
\end{\eq}
with the characteristic \temp ${T_+}'\equiv -T_+$. With this form of
the entropy the second law is guaranteed.
The entropy ${S_{W(\tot)}}'$ is an increasing function of the area
of the outer horizon, consistently with the Bekenstein's argument \cite{Beke:73}.\\

(e). $\hO<0$, $\he  > 1$: This system is effectively the same as the
case (c), and the same entropy ${S_{\new}}'$ of (\ref{S:new'}) and
${T_-}'$ of (\ref{T-'}) apply.\\

(f). $\hO<0$, $\he  <-1$: This is effectively the same system as
that of the case (b), and so the same entropy ${S_{\new}}$ of
(\ref{S:new}) and ${T_-}$ of (\ref{T-}) apply.\\

\begin{section}
{Inclusion of a gravitational Chern-Simons term (II): Statistical
entropy}
\end{section}

In the absence of the higher {\it curvature} terms, the central
charges of the holographic anomalies (\ref{anomaly}) are obtained
as
\begin{\eq}
\hat{c}^{\pm}_{GCS}=\gamma^{\pm} \frac{3 l}{2G \hbar},
\label{c:anomaly}
\end{\eq}
with $\gamma^{\pm}=1\pm {\hb}$ for the right/left-moving sectors,
respectively \cite{Krau:05a,Solo:05a}. On the other hand, in the
absence of the gravitational Chern-Simons term, the central charges
are given by (\ref{c:anomaly(high)}). Now when both the higher
curvature and gravitational Chern-Simons terms present, their
contributions are summed to obtain the total central charges as
follows \cite{Saho:06}:
\begin{\eq}
\hat{c}^{\pm}_{\tot}&=&\frac{3 l}{2G \hbar}+(\hO-1) \frac{3 l}{2G
\hbar}+ (\gamma^{\pm}-1) \frac{3 l}{2G \hbar} \no \\
&=&\hO (1\pm \he) \frac{3 l}{2G \hbar}. \label{c:anomaly(tot)}
\end{\eq}
And also, regarding the Virasoro generators and their ground state
generators $\hat{L}^{\pm}_{0(\tot)}$, they are also summed to get
the {\it total} Virasoro generators: In the absence of the higher
curvature terms, the ground state Virasoro generators, in the
standard form of a CFT on the plane, are given by
\begin{\eq}
\label{ground:GCS}
 \hat{L}^{\pm}_{0(GCS)}&=&\frac{lM_{GCS} \pm J_{GCS}}{2\hbar} +\f{
\hat{c}^{\pm}_{GCS}}{24
}\no \\
&=&\gamma^{\pm} \frac{(lm \pm j)}{2\hbar} +\f{
\hat{c}^{\pm}_{GCS}}{24 },
\end{\eq}
whereas in the absence of the gravitational Chern-Simons term,
$\hat{L}_0^{\pm}$ are given by $\hat{L}^{\pm}_0= \hO \frac{(lm \pm
j)}{2\hbar} +\f{ \hat{c}^{\pm}}{24 }$ as in (\ref{ground}). So, the
total ground state generators are given by
\begin{\eq}
\label{ground:tot}
 \hat{L}^{\pm}_{0(\tot)}&=&[1+(\hO-1)+(\hat{\ga}^{\pm}-1)] \frac{(lm \pm j)}{2\hbar} +\f{
\hat{c}^{\pm}_{\tot}}{24 } \no \\
 &=&\hO (1 \pm \he) \frac{(lm \pm j)}{2\hbar} +\f{
\hat{c}^{\pm}_{\tot}}{24 } \no \\
 &=&\frac{lM_{\tot} \pm J_{\tot}}{2\hbar} +\f{
\hat{c}^{\pm}_{\tot}}{24 }.
\end{\eq}

Now, with the above CFT data
$(\hat{c}^{\pm}_{0(\tot)},\hat{L}^{\pm}_{0(\tot)})$, one can compute
the statistical entropy for the asymptotic states, from the Cardy
formula (\ref{Cardy}) with the appropriate conditions (\ref{cond1})
and (\ref{cond2}), as follows [ $\hat{\ga}^{\pm} \equiv \hO (1 \pm
\he)$ ]
\begin{\eq}
S_{\stat(\tot)}&=&\mbox{log}~ \rho
(\hat{\Delta}^{+}_{\tot})+\mbox{log}~ \rho
(\hat{\Delta}^{-}_{\tot}) \no \\
&=&\f{\pi}{4 G \hbar} |\hO (1 +\he) (r_+ + r_-)|+\f{\pi}{4 G \hbar}
|\hO (1 -\he)
(r_+ - r_-)| \no \\
&=&\f{\pi}{4 G \hbar} (|\hat{\ga}^+| +|\hat{\ga}^-|) r_+ +\f{\pi}{4
G \hbar} (|\hat{\ga}^+|-|\hat{\ga}^-|)r_-~,
\label{S:stat(tot)}
\end{\eq}
where $\hat{\Delta}^{+}_{\tot}$ are the eigenvalues of the operators
$\hat{L}^{\pm}_{0(\tot)}$ for black-hole quantum states
$|\hat{\Delta}^{\pm}_{\tot}\rangle$ and I have chosen their minimum
values as $\hat{\Delta}^{\pm}_{ \hbox{\scriptsize min} (\tot)}=0$;
from (\ref{ground:tot}), this corresponds to the $AdS_3$ vacuum
having $m=-1/(8G)$ and $ j=0$ in the usual context as usual, but it
has a permanent rotation, as well as the conformal-factor
corrections,
\begin{\eq}
M_{\tot}=-\f{\hO}{8G},~~J_{\tot}=-\f{l \hO \he}{8G}
\end{\eq}
in the new context \cite{Krau:05a}.

Then, let me consider the following four cases, depending on the
values of $\hO$ and $\he$: (a). $\hO>0$, $|\he | \leq 1$, (b).
$\hO>0$, $\he > 1$ or $\hO<0$, $\he < -1$, (c). $\hO>0$, $\he < -1$
or $\hO<0$, $\he > 1$, (d). $\hO<0$, $|\he | \leq 1$.\\

(a). $\hO>0$, $|\he | \leq 1$: In this case, I have
$|\hat{\ga}^{\pm}|=\hat{\ga}^{\pm}$ and the statistical entropy
(\ref{S:stat(tot)}) becomes
\begin{eqnarray}
\label{S_stat:old2}
 S_{\stat (\tot)}=\hO \left(~\frac{2 \pi r_+}{4 G \hbar}+\he \frac{2 \pi r_-}{4
G \hbar} ~\right)
\end{eqnarray}
from $\hat{\ga}^++\hat{\ga}^-=2 \hO,~\ga^+-\ga^-=2 \hO \he$. This
agrees exactly with the usual entropy formula (\ref{S:Wald(tot)}),
which agrees with the Wald's formula also \cite{Saho:06}. And, this
is the case where $\hat{c}^{\pm}_{\tot}$ and
$\hat{\Delta}^{\pm}_{\tot} -\hat{c}^{\pm}_{\tot}/{24}$ are positive
definite such as the Cardy formula (\ref{Cardy}) has a well-defined
meaning. In the gravity side also it shows the usual behavior with
the ``positive'' mass and angular momentum, satisfying
the normal inequality $M^2_{\tot} \ge J^2_{\tot}/l^2$.\\

(b). $\hO>0$, $\he > 1$ or $\hO<0$, $\he < -1$: In this case, I have
$|\hat{\ga}^{+}|=\hat{\ga}^{+},~|\hat{\ga}^{-}|=-\hat{\ga}^{-}$ and
so the statistical entropy (\ref{S:stat(tot)}) becomes
\begin{eqnarray}
\label{S_stat:new}
 S_{\stat (\tot)}=\hO \left(~\frac{2 \pi r_-}{4 G \hbar}+\he \frac{2 \pi r_+}{4
G \hbar} ~\right).
\end{eqnarray}
This agrees exactly with the new entropy formula (\ref{S:new}),
which guarantees the second law of thermodynamics in this case. And,
this is the case where there is an abnormal change of the role of
the mass and angular momentum due to $J_{\tot}^2/l^2\ge M_{\tot}^2
$, even though $M_{\tot}$ and $J_{\tot}$ both are positive definite,
as usual. Moreover, in the CFT side also, this is not the usual
system either because $\hat{c}^{-}_{\tot}=\hat{\ga}^- 3 l/(2G
\hbar)$ and $\hat{\Delta}^{-}_{\tot}
-\hat{c}^{-}_{\tot}/{24}=\hat{\ga}^-_{\tot} (ml -j)/2\hbar$ are
$negative$ valued, though their self-compensations of the negative
signs produce the $real$ and $positive$ statistical entropy.
However, the CFT is perfectly well defined, as I have discussed in
the case (ii) of Sec. III, by considering another representation of
the Virasoro algebra with $\tilde{L}_{n(\tot)}^-\equiv
-\hat{L}_{-n(\tot)}^-,~\tilde{c}^-_{\tot}\equiv-\hat{c}^-_{\tot}$
and $\tilde{L}^-_{n(\tot)}|\tilde{\De}^-_{\tot}
\left.\right>=0~(n>0)$ for a new highest-weight state
$|\tilde{\De}^-_{\tot}\left.\right>$ \cite{Park:06,Park:06b}; this
implies that the Hilbert space need to be ``twisted'' in which the
whole states vectors be constructed from the twisted highest-weight
state $\left.|\hat{\Delta}^{+}_{\tot}\right>
\otimes\left.|\tilde{\De}^-_{\tot}\right>$, in contrast to the {\it
double} twistings for the case (ii) of Sec. III. The formula
(\ref{S:stat(tot)}), which is invariant under this substitution,
should be understood in this context. On the other hand,
 if I take the limit $\he \ra \infty$, this becomes the ``exotic'' black hole
 system that occurs in several different contexts
\cite{Carl:91,Carl:95,Bana:98a,Bana:98c,Park:06}. Interestingly,
this includes the limiting case of $\hO \ra 0$ with {\it any finite,
non-vanishing} $\hb =\hO \he$,  as well as the case of $\hb \ra
\infty$, with the finite $\hO$, in which there is only the
 gravitational Chern-Simons term, without the Einstein-Hilbert and
 its higher curvature corrections, \\

(c). $\hO>0$, $\he < -1$ or $\hO<0$, $\he > 1$: In this case, I have
$|\hat{\ga}^{+}|=-\hat{\ga}^{+},~|\hat{\ga}^{-}|=\hat{\ga}^{-}$ and
the statistical entropy (\ref{S:stat(tot)}) becomes
\begin{eqnarray}
\label{S_stat:new2}
 S_{\stat(\tot)}=-\hO \left(~\frac{2 \pi r_-}{4 G \hbar}+\he \frac{2 \pi r_+}{4
G \hbar}~\right).
\end{eqnarray}
Note that this is positive definite and this should be the case due
to the definition
$S_{\stat(\tot)}=\log(\rho(\hat{\De}^+_{\tot})\rho(\hat{\De}^-_{\tot}))\ge
0$ for the number of states $\rho (\hat{\De}^{\pm}_{\tot}) \geq 1$.
This agrees exactly with the {\it modified} new entropy formula
(\ref{S:new'}), which guarantees the second law. And this is the
case where $M_{\tot}$ $can$ be negative and $J_{\tot}$ has the
opposite direction to the bare one $j$, in contrast to the positive
definite $M_{\tot}$ and $J_{\tot}$ in the cases of (a) and (b), as
well as the anomalous inequality $J_{\tot}^2/l^2 \geq M_{\tot}^2$.
In the CFT side, $\hat{c}^{+}_{\tot}$ and $\hat{\Delta}^{+}_{\tot}
-\hat{c}^{+}_{\tot}/{24}$ become negative-valued now, and I need to
twist this right-moving sector, rather than the left-moving one as
in the case (b), $\tilde{L}_{n(\tot)}^+\equiv
-\hat{L}_{-n(\tot)}^+,~\tilde{c}^+_{\tot}\equiv-\hat{c}^+_{\tot}$
and $\tilde{L}_{n(\tot)}^+|\tilde{\De}^+_{\tot}
\left.\right>=0~(n>0)$ for the twisted highest-weight state
$\left.|\tilde{\Delta}^{+}_{\tot}\right>
\otimes\left.|\hat{\De}^-_{\tot}\right>$.
\\

(d). $\hO<0$, $|\he | \leq 1$: In this case, I have
$|\hat{\ga}^{\pm}|=-\hat{\ga}^{\pm}$ and the statistical entropy
(\ref{S:stat(tot)}) becomes
\begin{eqnarray}
\label{S_stat:old2}
 S_{\stat (\tot)}=-\hO \left(~\frac{2 \pi r_+}{4 G \hbar}+\he \frac{2 \pi r_-}{4
G \hbar} ~\right).
\end{eqnarray}
This agrees exactly with the modified entropy formula
(\ref{S:W(tot)'}), which is positive definite and guarantees the
second law as well as the first law. This case is exactly the same
situation as in the case (ii) of Sec. III, and there is the upper
bound for the mass, i.e., $M_{\tot} \leq J_{\tot}/l \leq 0$, though
one has the usual inequality in the magnitudes $M^2_{\tot} \geq
J^2_{\tot}/l^2$. And, in contrast to the above (b), (c) cases, I
need the {\it doubly}-twisted highest-weight state
$|\tilde{\De}^+_{\tot}\left.\right> \otimes
|\tilde{\De}^+_{\tot}\left.\right>$ with the new presentation of the
Virasoro algebras for $\tilde{L}_{n(\tot)}^\pm\equiv
-\hat{L}_{-n(\tot)}^\pm,~\tilde{c}^\pm_{\tot}\equiv-\hat{c}^\pm_{\tot}$,
and $\tilde{L}_{n(\tot)}^\pm|\tilde{\De}^\pm_{\tot}
\left.\right>=0~(n>0)$. It is interesting to note that I have
perfectly well-defined CFT for all the possible cases and {\it there
are no conflicts, in the general theory of (\ref{I:tot}), between
the ``single'' twisting of the Hilbert space with the gravitational
Chern-Simons of Sec.III and the ``double'' twistings with the higher
curvatures \cite{Park:06,Park:06b} }: They are well self-organized
such as the negative norm states do not occur.\\

In summary, I have found exact agreements between the new
thermodynamical black hole entropies for the bulk (AdS) gravity with
the gravitational Chern-Simons as well as the general higher
curvature terms and the CFT entropies in the asymptotic boundary,
for $all$ values of the coupling $\he$ and the conformal factor
$\hO$. It is remarkable that CFT has no conflict, for the general
theory with both the higher curvature and gravitational Chern-Simons
terms, between the single twisting of the Hilbert space for the
gravitational Chern-Simons term and double twisting for the higher
curvatures such as the CFT can be well-defined for all cases, though
this is not so clear at first. So, the new entropy formulae for the
strong coupling $|\he|
>1$, either $\hO>0$ or $\hO<0$
and the modified entropy formula for $\hO <0, |\he| \leq 1$
seem to be supported by the CFT approach also.
This reveals the AdS/CFT correspondence in the sub-leading orders
with the ``all'' higher curvature terms and the higher derivative
term of the gravitational Chern-Simons, as well as in the leading
order
with the Einstein-Hilbert action.\\

\begin{section}
{Summary and open problems}
\end{section}

I have studied the thermodynamics of the BTZ black hole in the
presence of $all$ the higher curvature terms and a gravitational
Chern-Simons term, and its solid connection with a statistical
approach, based on the holographic anomalies.

The main results are as follows:

First, for the case of a large coupling $|\he| >1$, with any value
of $\hO$, the new entropy formulae
are proposed, from the purely thermodynamic point of view such as
the second law of thermodynamics be guaranteed.

Second, for the case of $\hO <0, |\he| \leq 1$, the modified
(Wald's) formula
is proposed from purely the second law.

Third, I have found supports for the proposals from a CFT based
approach which reproduces the new entropy formulae for $|\he|>1$ and
the modified formulae for $\hO <0, |\he| \leq 1$, as well as the
usual entropy formula for a small coupling $|\he| \le 1$, $\hO>0$.
This would provide a non-trivial check of the AdS/CFT
correspondence, in the presence of higher curvature/derivative terms
in the gravity theory. I have also found that there is no conflict,
for the general theory with both the higher curvature and
gravitational Chern-Simons terms, from the different Hilbert space
for each term.

Some open problems would be the followings:

1. A difficult problem of the new entropy formulae is that they
require rather unusual characteristic temperature $T_-=\kappa/(2
\pi)|_{r_-}$ or \temp ${T_+}'=-T_+$,
being {\it negative}-valued, or ${T_-}'=-T_-$, and angular velocity
$\Om_-$, being the inner-horizon angular velocity in the BTZ black
hole. The ``negative'' temperature is quite well defined in the
statistical mechanics when there is an upper bound of the energy
levels \cite{Kitt:67}. This situation is is quite similar to our
case where the entropy is a function of the mass and there is upper
bound of mass. So, the negative-valued temperature {\it might} not
be so strange in this context.
But this is contrast to the Hawking
temperature in the usual Hawking radiation whose radiation spectrum
is determined by the metric alone \ci{Hawk:75}.



2. Can we compute the ``classical'' Virasoro algebra ``directly'' in
the higher curvature frame, without recourse to the frame
transformation to the theories without the higher curvatures ?

$2'$. Can we explicitly prove that the auxiliary tensor matters
which appear after the frame transformation have ``no''
contributions to the Virasoro algebras, such as there are perfect
agreements between the holographic anomalies and the classical
Virasoro algebra approaches ? This would provide a more ``direct''
check of the AdS/CFT correspondence.

3. Can we find the gauge theoretic formulations of the higher
curvature gravities ? This would provide a more ``explicit''
computation of the classical Virasoro algebras
\cite{Park:98,Park:04a,Park:06b}.

Complete answers to these problems are still missing. But, as far as
the first open problem is concerned, I have recently proposed how
this might be circumvented by noting some possible limitations of
the standard approach initiated by Hawking \ci{Hawk:75} in our
unusual circumstances \ci{Park:06,Park:0610}.
Here, the important
point would be that a {\it dynamical} geometry responds differently
under the emission of Hawking radiation, even though the formal
metric is the same
\ci{Park:06}. For
the case of {\it negative} conformal factor ($\hO<0$) in the higher
curvature black hole in Sec. II, for example, the emission of a
particle with a (positive) energy $\omega$ would reduce the black
holes's mass $M$, which being negative valued, from the conservation
of energy, but this corresponds to the ``increasing'' of the
(positive) mass $m$ in the ordinary BTZ black hole context, due to
the negative factor in (2.8). So, this implies that the horizon, in
the ordinary BTZ black hole context, expands as it emits Hawking
radiation with a positive energy, in contrast to the case of
positive mass black hole.

Here, the conservations of energy and angular momentum, which are
not well enforced in the standard computation, would have a crucial
role. In this respect, the Parikh and Wilczek's approach
\ci{Pari:00}, which provides a direct derivation of Hawking
radiation as a quantum tunneling by considering the global
conservation law naturally, would be an appropriate framework to
study this problem. This is currently under study.\\


\section*{Acknowledgments}

I would like to thank Jacob Bekenstein, Gungwon Kang, Sergei
Odintsov, and Ho-Ung Yee for useful correspondences. This work was
supported by the Korea Research Foundation Grant funded by Korea
Government(MOEHRD) (KRF-2007-359-C00011).

\newcommand{\J}[4]{#1 {\bf #2} #3 (#4)}
\newcommand{\andJ}[3]{{\bf #1} (#2) #3}
\newcommand{\AP}{Ann. Phys. (N.Y.)}
\newcommand{\MPL}{Mod. Phys. Lett.}
\newcommand{\NP}{Nucl. Phys.}
\newcommand{\PL}{Phys. Lett. }
\newcommand{\PR}{Phys. Rev. }
\newcommand{\PRL}{Phys. Rev. Lett.}
\newcommand{\PTP}{Prog. Theor. Phys.}
\newcommand{\CQG}{Class. Quant, Grav.}
\newcommand{\hep}[1]{ hep-th/{#1}}
\newcommand{\hepg}[1]{ gr-qc/{#1}}
\newcommand{\bi}{ \bibitem}

\end{document}